# Molecule-induced surface second-order nonlinearity in an inversion symmetric microcavity

Ru Wang[#], Yue Dai[#], Jinsong Cheng, Ruoyu Wang, Xiaoqin Shen*

School of Physical Science and Technology, ShanghaiTech University, Shanghai, China 201210

[#] The authors contributed equally to this work.

Corresponding email: shenxq@shanghaitech.edu.cn

**Abstract**: Inversion symmetry eliminates the second-order nonlinear responses in materials commonly used in silicon photonics with electric-dipole approximation. The lack of effective methods to induce the second-order nonlinearity in silicon photonic materials prevents their applications in second-order nonlinear integrated photonics. Here, we experimentally demonstrate a surface second-order nonlinear optics approach for boosting the second harmonic (SH) generation process in a silica microcavity. By leveraging the molecule-induced surface second-order nonlinearity, a record high SH efficiency of about 6.7% $W^{-1}$ is achieved in a silica microcavity functionalized with a surface asymmetrically-aligned molecular monolayer, which is enhanced of two to four orders of magnitude compared to that before molecule-functionalization. Furthermore, we derive the equations that govern the surface second-order nonlinear process in inversion symmetric microcavities. Our method not only enables high efficiency second-order nonlinear frequency conversions in silica photonics, but also can apply to other inversion symmetric material platforms for integrated photonics.

Second-order nonlinearity lies at the core of modern photonics for photonic applications in both classical and quantum regimes [1-3]. It is of paramount importance to realize integrated frequency microcombs [4-9], quantum light generation [10, 11], and frequency self-referencing [12, 13]. However, for typical silicon photonics materials (silica, silicon and silicon nitride), inversion symmetry eliminates the bulk second-order nonlinear response within the electric dipole approximation [14]. Alternatively, asymmetric materials such as lithium niobate [15, 16], aluminum nitride[17], gallium arsenide [18, 19], and silicon carbide [20, 21], are intensively developed in recent years with the advances in nano-fabrication techniques [22]. Yet, they are still of high fabrication complexity and lack of compatibility with silicon-based complementary metal-oxide semiconductor technology. Therefore, there is a persisting interest in developing strategies for inducing high second-order nonlinearity in silicon photonic materials and devices.

Early efforts involved in externally applying of strains [23, 24] or electric fields [25] to waveguides for inducing second-order nonlinear responses in silicon photonic materials. Recently, by leveraging the resonance enhancement effect in high quality factor (Q) whispering gallery mode (WGM) microcavities, SH signals are demonstrated in silica [26] and silicon nitride [27] microcavities with conversion efficiency of 0.049% W$^{-1}$ and 0.1% W$^{-1}$, respectively. The SH efficiencies are limited by the weak intrinsic $\chi^{(2)}$ induced by spontaneous surface symmetry breaking. By further utilizing the photo-induced photogalvanic effect or optical-poling effect, the nonlinearities are further improved in silicon nitride microcavities [28, 29], leading to the enhancement in SH efficiencies. However, the photo-induced nonlinearity is fundamentally limited by the weak intrinsic $\chi^{(2)}$. In addition, it exhibits a slow time response and high pump power requirement that could be an obstacle form some applications. Alternatively, there is an increasing interest in surface molecule approach for the second-order nonlinearity in inversion symmetric microcavity devices [30, 31]. These studies, however, did not consider either resonance phase-matching condition or the effective surface molecular nonlinearity. In addition, the theoretical model for surface second-order nonlinear process in WGM microcavity is still vague [30]. Thus, challenges remain to explore the molecule-induced surface second-order nonlinearity in inversion symmetric microcavities for integrated photonics.

In this letter, we demonstrate, for the first time, the molecule-induced surface second-order nonlinearity in a WGM microcavity for boosting the second harmonic (SH) conversion efficiency. We use a high quality factor silica microsphere cavity for the demonstration. The silica microcavity is functionalized with an asymmetrically-aligned monolayer of nonlinear molecule on the surface, resulting in a strong surface molecular second-order nonlinearity (**Fig. 1a**). Under dual mode resonance phase matching (also referred as perfect phase matching) condition, the device generates strong SH signal (~775 nm) under a continuous wave pump laser (~1550 nm) with a few miliwatt power. With "in situ" measurement, we reveal that the SH efficiency is improved by two to four orders of magnitude, which is attributed to the monolayer molecules interacting with the optical evanescent field at the microcavity surface. Moreover, we analyze the theoretical mode that takes account of the surface interaction between the evanescence field with the molecule monolayer and derive the equations for the molecule-induced surface second-order nonlinear process. It indicates that the SH efficiency could be remarkably boosted up to nine to ten orders of magnitude with optimized geometry design. The demonstrated results open a door towards high efficiency second-order nonlinearity in integrated silicon photonics.

By analyzing the theoretical model for surface second-order nonlinear optics in a whispering gallery mode (WGM) microcavity (see Supplemental Materials [32], Sec. 1). The overall second-order nonlinear electric polarization can be expressed as: $\boldsymbol{P}^{(2)} = \chi^{(2)}_{silica} : \boldsymbol{E_1 E_1} + \rho^2 \chi^{(2)}_{mol} : \boldsymbol{E_1 E_1}$, where $\chi^{(2)}_{silica}$ the apparent second-order nonlinear susceptibility of the microcavity material (silica), $\chi^{(2)}_{mol}$ second-order nonlinear susceptibility originates from the strong molecular electric-dipole response, $\rho$ is the intensity ratio of the evanescent field to the peak electric field of the pump mode (**Fig. 1a-b**). Under dual resonance phase matching and critical coupling, the SH efficiency derived from the coupled mode equation can be expressed as (see Supplemental Materials [32] Sec. 2-3):

$$P_2/P_1^2 = \frac{4Q_2}{\omega_2}\left(\frac{4Q_1}{\omega_1}\right)^2 K^2 \left[\chi^{(2)}_{silica} + \rho^2 \chi^{(2)}_{mol}\right]^2, \quad (1)$$

where the subscripts j = 1, 2 represent the pump and SH cavity modes with resonance frequencies $\omega_j$, respectively; $Q_j$ is the loaded quality factors, respectively; $P_1$ are the pump power; $K$ is defined as the coupling factor to get $K\chi^{(2)}$ the second-order nonlinear coupling strength between the two modes. While multiple factors, such as Q values, critical coupling and phase matching, can be optimized for SH generation in a silica microcavity, the conversion efficiency is fundamentally determined by the second-order nonlinearity. As can be found in equation 1, the efficiency is quadratically dependent on the effective second-order nonlinear susceptibility ($\chi^{(2)}_{silica} + \rho^2 \chi^{(2)}_{mol}$).

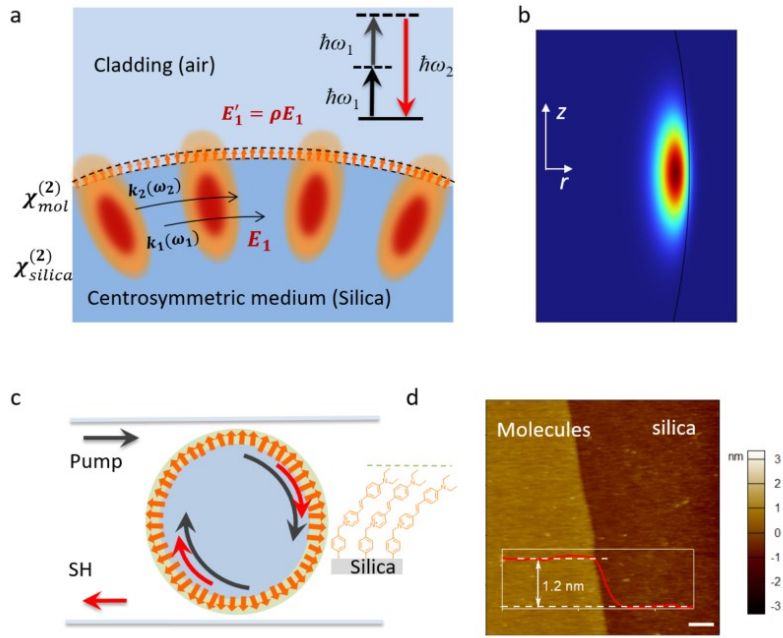

**Fig. 1 Molecule-enhanced microcavity SHG. a**, Schematic of the phenomenological model for SHG at the surface of a whispering gallery mode microcavity made of symmetry materials. The effective intensity of evanescent field ($\rho E_1$) is taken into account of the surface nonlinear optical process. **b**, FEM simulation of the fundamental optical mode profile (at 1550 nm) in the microcavity with a radius of 30 μm. $z$ is the polar axis and $r$ is the radical axis. The black line indicates the boundary between the silica device and the cladding (air). The optical field extended in air is about 1.7% of the total optical field ($\rho = 0.017$). **c**, Schematic of SHG from a surface functionalized microcavity. A continuous wave pump laser is coupled into the microcavity via a tapered optical fiber. A secondary tapered fiber is used to couple out the generated SH signal. Inset is the schematic of the surface monolayer of the molecules (denoted as orange arrows in **a** and **c**) **d**, AFM mapping of a silica surface functionalized with a layer of stibazolium molecules. Inset is the measured step height of the molecular layer. Scale bar: 1 μm.

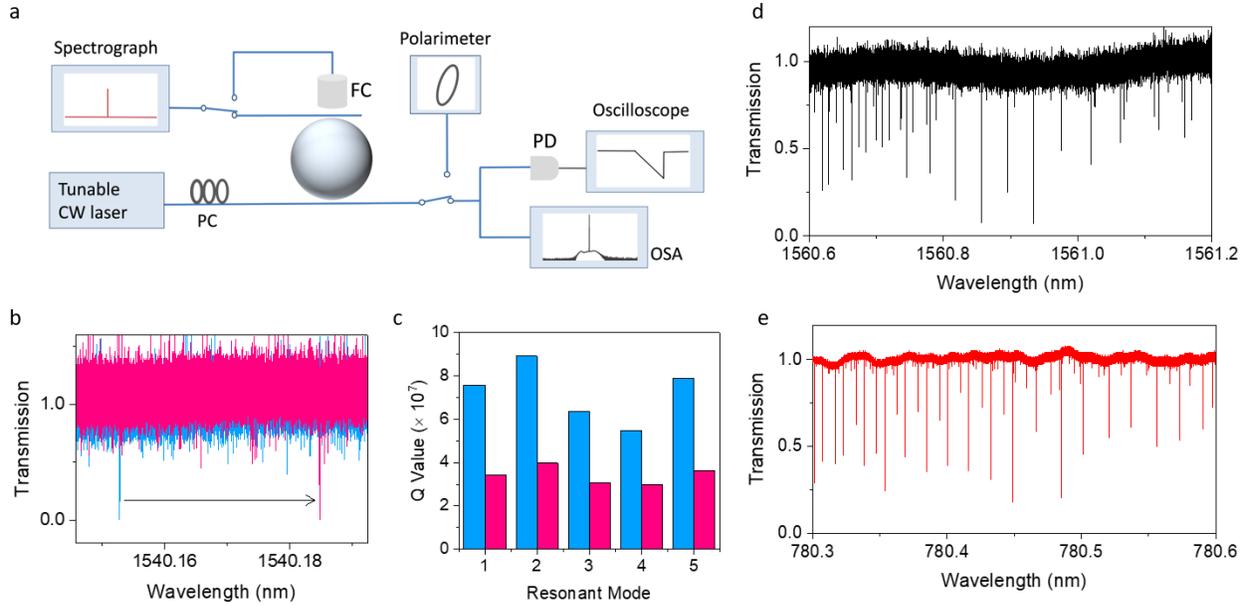

**Fig. 2 Microcavity optical characterization. a**, Optical set-up. PC, polarization controller; PD, photo detector; OSA, optical spectrum analyzer; FC, fiber collimator. The generated SH signal is either coupled out with a signal optical tapered fibre or detected with a FC, and then is sent to a spectrograph. **b**, Transmission spectra at ~ 1550 nm of a silica microcavity before and after surface molecule functionalization. The resonance peak red shifts of about 30 pm after surface functionalization. The laser power is set bellow 1 μW to avoid any opto-thermal effect. **c**, The measured intrinsic Q values of six different modes of a silica microcavity before and after surface functionalization. The functionalized silica microcavity exhibit high Q values of about $3 \times 10^7$, about one half of that of the freshly prepared bare silica microcavity before functionalization. **d-e**, The representative transmission spectra of a functionalized silica microcavity for the pump modes at ~ 1550 nm (**d**) and the SH modes ~780 nm (**e**).

The silica microcavity is tested by using a continuous wave (CW) laser. Light form a tunable laser is coupled into the cavity via a single-mode tapered fibre waveguide (**Fig. 2a**). The resonance wavelength at ~1550 nm of the surface functionalized silica microcavities red shifts by about 30 pm, compared to the silica microcavities before functionalized (**Fig. 2b**). It can be ascribed to the slightly increase in the cavity diameter and the effective refractive index. The functionalized silica microcavities exhibit high Q values of about $3 \times 10^7$ for both SH (~765 nm) and pump (~1550 nm) cavity modes, slightly lower than that of the freshly prepared bare silica microcavities (**Fig. 2c**). As can be shown in **Figure 2d-e**, the cold pump modes of SH and pump are typically phase-mismatched in terms of frequency due to the dispersion.

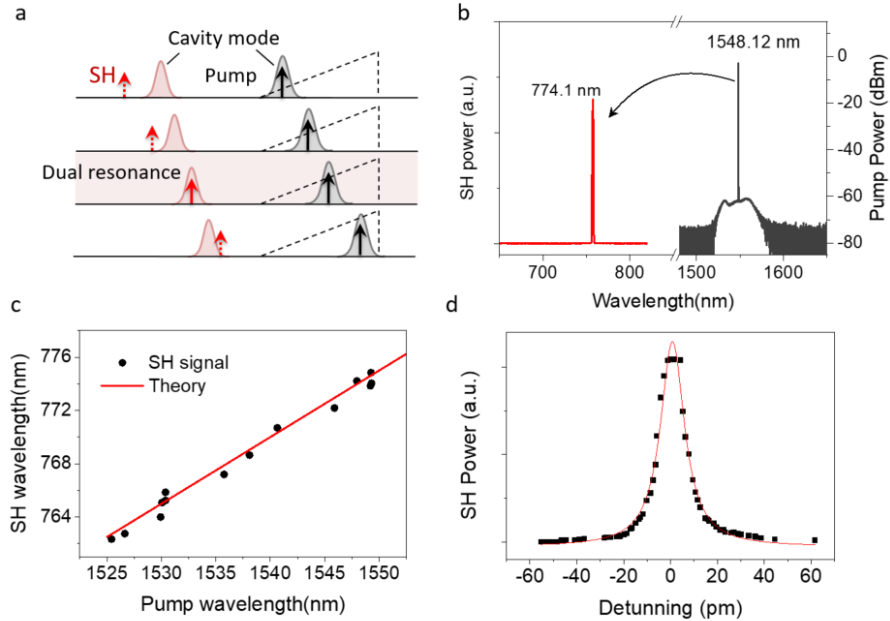

**Fig. 3 Dual resonance phase matching for SH generation in the microcavities. a,** Schematic of the dual resonance phase matching process. The back and red arrows denote the pump and SH wavelength, respectively. The grey and red Lorentzian peaks denote the pump and SH cavity mode, respectively. The dashed black triangles indicates the thermo-optic bistability in the high Q cavity. The x axis represents the wavelength and SH generation can happen when both the pump and SH light on resonance with their cavity modes. **b**, A representative spectrum of the SH signal and its corresponding pump. **c**, SH wavelengths versus the corresponding pump wavelengths for different modes. **d**, SH power versus pump detuning with input power of about 4.5 mW.

To generate SH, we leverage the cavity-enhanced thermal and optical Kerr effects to realize dual resonance phase-matching condition[26] (**Fig 3a**). The pump laser is scanned from blue detuning to red detuning of a cavity mode to couple power into the cavity. At this moment, only the pump light is on resonance. Owing to the cavity-enhanced thermal and optical Kerr effects, both the pump and SH modes shift consequently to longer wavelengths. By increasing the wavelength of the pump light to match the resonant mode, at a certain pump wavelength, the SH signal can also match the SH cavity mode. Therefore, the dual resonance phase matching condition is achieved. **Figure 3b** shows a representative SH spectrum and the corresponding pump spectrum measured by a spectrograph coupled with electron-multiplying charge-coupled device (EMCCD) and an optical spectrum analyzer, respectively. The SH signal appears at 774.1 nm when the device is pumped at 1548.12 nm. Due to the existence of a rich variety of cavity modes at ~765 nm (SH) and ~1550 nm (pump) ranges, SH signals can be generated under different pump wavelengths from 1510 nm to 1560 nm (**Fig 3c**). The maximum SH signal can be generated by carefully tuning of the pump frequency (**Fig 3d**). For the functionalized microcavity, it exhibits high conversion efficiency $P_2/P_1^2$ of 4.7% W$^{-1}$. To our best knowledge, it is the highest SH efficiency reported for silica devices.

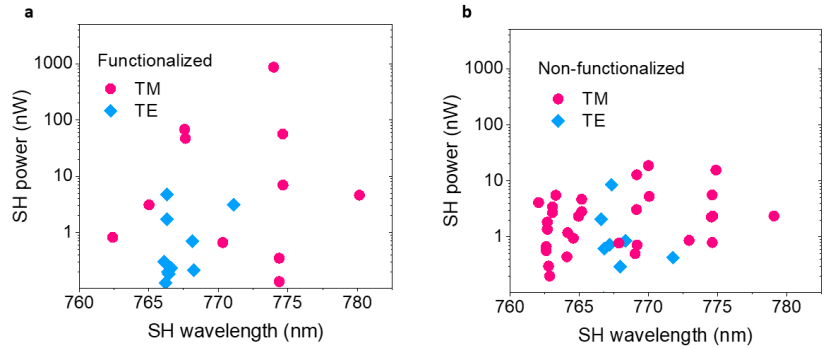

**Fig. 4 Polarization dependency of SH generation in a functionalized (a) and a non-functionalized (b) silica microcavity.** The maximum SH signal with a TM pump is two orders of magnitude (~180 times) higher than that with a TE pump in the functionalized silica microcavity (**a**). The maximum SH signal with a TM pump is only a few times higher than that with a TE pump in the non-functionalized silica microcavity (**b**). The pump power is fixed at ~ 5.2 mW, and the coupling conditions are optimized for TM and TE polarization.

We verify the dependence of the SH signal on the polarization of the pump light. The effective $\chi_{mol,\perp}^{(2)}$ tensor in the radical direction of the microsphere would contribute to the surface SH process under a transverse magnetic (TM) pump. Experimentally, the SH signals under optimal coupling condition and a fixed pump power are recorded by pumping different modes with pre-identified polarization. **Figures 4** compares the pump polarization dependence of SH conversion for the surface functionalized and non-functionalized silica microsphere cavities, respectively. It is found that the maximum SH signal with TM pump is two orders of magnitude (~180 times) higher than that with transverse electric (TE) pump in the surface functionalized microcavity. As the symmetrically-aligned surface molecules tend to yield a net dipole moment along the normal direction of the microsphere surface, the nonlinear enhancement happens under TM pump modes. In contrast, the maximum SH signal with a TM pump is only a few times higher than that with a TE pump in the non-functionalized microcavity [26]. In addition, for both the functionalized and non-functionalized microcavity, the maximum SH signals under TE pump has no obvious difference, indicating the randomness of the tilting direction of the surface molecules. Besides, **Figure 4** also shows that the SH signal under different pump modes (both TM and TE) varies significantly, which can be mainly attributed to the different pump-SH mode overlap for different pump modes.

In additional to phase matching, several complex factors affect the SH efficiency (as implied in equation 1), including the device's size, Q factor, coupling, pump-SH mode overlap, as well as fabrication defects and bulk inhomogeneity that could induce a sensitive SH response. To more quantitatively analyze the surface molecule-induced SH enhancement capability, we carefully carry out an "in situ" experiment (**Fig. 5a**). We use a silica microcavity device with a same taper fiber to measure the SH power under same coupling conditions before and after surface molecular functionalization. The SH power is measured with a TM pump under optimal fiber coupling and a fixed pump power. Thus, the influences caused by the variations in diameter, defects and bulk in homogeneity in different devices can be excluded. Besides, by comparing the SH signals under the same pump mode, the influence of the variation in mode overlap could further be minimized. **Figure 5b** shows that, for the tested device before and after surface molecular functionalization, profound SH signals appear under different pump modes with spacing of about one FSR (from1510 nm to 1535 nm). The maximum SH signals before and after molecular functionalization are both found under the TM pump at about 1521.45 nm. By direct comparing the SH signals under the same pump mode, it can be found that the SH power is improved by about 27 times. The surface molecule-induced SH enhancement factor is derived from the experimental data to be 216, which agrees well the theoretical value.

Furthermore, we quantitatively analyze the SH enhancement factor contributed from the interaction between the surface molecules and the evanescent field. Given that the Q factor and other factors remain constant, the enhancement factor under a TM pump can be expressed as $\eta_{mol} = (1 + \rho^2 \chi^{(2)}_{mol,\perp}/\chi^{(2)}_{silica})^2$, where $\chi^{(2)}_{silica}$ is about $10^{-3}$ pm/V for silica [33]. Remarkably, as shown in **Figure 5c**, by increasing the $\rho$ value from 0.017 (in our case) to 0.05 via proper geometry design, the surface molecule-induced SH enhancement factor can be improved to four-orders of magnitude. In addition, the $\chi^{(2)}_{mol,\perp}$ can be further improved by optimizing the molecular monolayer with strong nonlinear molecules, to further improve the SH enhancement factor by about one order of magnitude. Moreover, the maximum the molecule-induced SH enhancement factor can be expected to be about nine to ten orders of magnitude (**Fig. 5d**), given the case that a molecular monolayer sheet is embed inside the microcavity structure to interact with the peak optical field (see Supplemental Materials [32] Sec. 4). These results suggest that the molecule-induced second-order nonlinearity method is particularly promising for higher efficiency integrated second-order photonics.

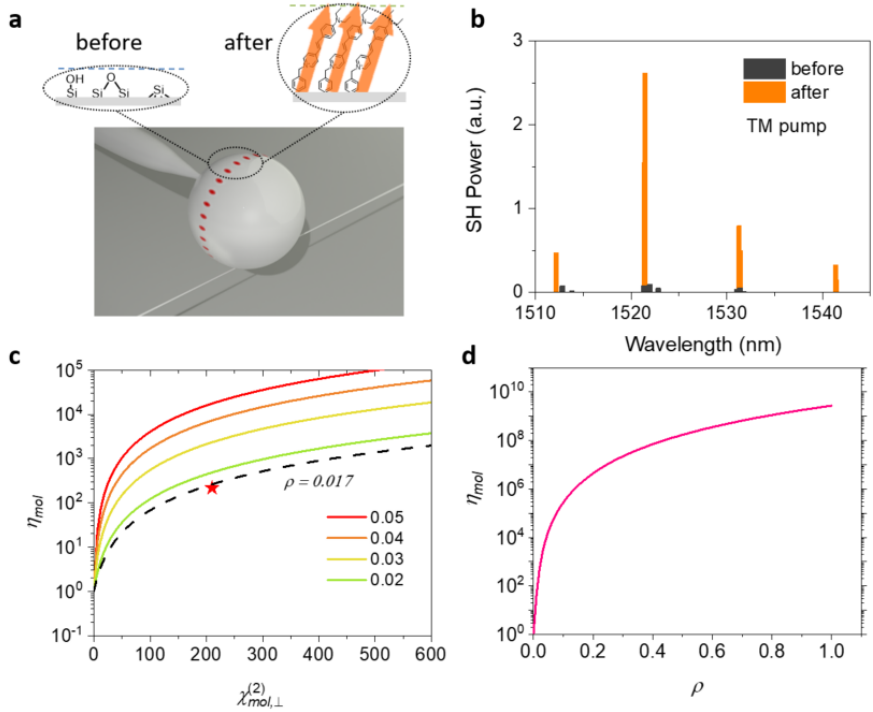

**Fig. 5 Surface molecule-enhanced second-order nonlinearity. a,** Schematic of the in-situ testing. A silica microcavity is tested under the same experimental condition before and after surface functionalization. TM pump modes in the microcavity is illustrated. **b**, The observed SH signals under different TM pump modes (with spacing of about one free spectrum range) in the silica microcavity before and after surface functionalization. **c**, The dependence of SH enhancement factor on the surface molecular nonlinearity $\chi^{(2)}_{mol,\perp}$. The red star and the grey dashed line represent the experimental data obtained from the functionalized silica microcavity in our work and the theoretical dependence of the enhancement factor on the $\chi^{(2)}_{mol,\perp}$ value of the surface molecular monolayer ($\rho = 0.017$). **d**, The dependence of SH enhancement factor on $\rho$.

In summary, we have demonstrated a molecule-induced surface nonlinear optics approach for boosting the second-order nonlinearity in silica microcavities. To verify our approach, we functionalized silica micocavities with a monolayer of nonlinear molecules and generate SH signals under dual mode resonance phase matching condition. Moreover, we analyze the theoretical mode and drive the equations for the surface nonlinear optical process that takes account of the effective molecular nonlinearity. We confirmed that the monolayer molecules interacting with the optical evanescent field at the microcavity would induce a strong intrinsic surface second-order nonlinearity, and enhance the SH efficiency by two to four orders of magnitude. This approach is not limited to the specific optical microcavity materials used in the present work, but highly implementable other inversion symmetric material platforms. The demonstrated results open up a promising avenue for high efficiency integrated second-order and hybrid nonlinear photonics [11, 12, 34].

**Acknowledgements**

We thank Yuan-Ron Shen for fruitful discussions. This project is supported by the National Natural Science Foundation of China (Grant No. 62275152), the Shanghai Pujiang Program (Grant No. 20PJ1411600) and the Science and Technology Commission of Shanghai Municipality (Grant No. 20ZR1436400). We also acknowledge the support from ShanghaiTech University, the Soft Mater Nanofabrication Laboratory.

**Author contribution**

R. Wang and Y. Dai conducted the fabrication and optical measurement. J. Cheng conducted the surface characterization. R. Wang conducted the FEM simulation. X. Shen conceived and supervised the project. X. Shen wrote the manuscript with inputs and comments from all the authors. R. Wang and Y. Dai contributed equally.

The authors declare no competing or conflicts of interest. Correspondence and requests for materials should be addressed to Xiaoqin Shen (shenxq@shanghaitech.edu.cn).

**References**

[1] R.W. Boyd, A.L. Gaeta, E. Giese, Nonlinear optics, Nonlinear optics, Springer2008, pp. 1097-1110.
[2] D.E. Chang, V. Vuletić, M.D. Lukin, Quantum nonlinear optics—photon by photon, Nature Photonics, 8 (2014) 685-694.
[3] C. Wang, Z. Fu, W. Mao, J. Qie, A.D. Stone, L. Yang, Non-Hermitian optics and photonics: from classical to quantum, Advances in Optics and Photonics, 15 (2023) 442-523.
[4] H.-J. Chen, Q.-X. Ji, H. Wang, Q.-F. Yang, Q.-T. Cao, Q. Gong, X. Yi, Y.-F. Xiao, Chaos-assisted two-octave-spanning microcombs, Nature communications, 11 (2020) 2336.
[5] S. Miller, K. Luke, Y. Okawachi, J. Cardenas, A.L. Gaeta, M. Lipson, On-chip frequency comb generation at visible wavelengths via simultaneous second-and third-order optical nonlinearities, Optics Express, 22 (2014) 26517-26525.
[6] A. Roy, L. Ledezma, L. Costa, R. Gray, R. Sekine, Q. Guo, M. Liu, R.M. Briggs, A. Marandi, Visible-to-mid-IR tunable frequency comb in nanophotonics, Nature Communications, 14 (2023) 6549.
[7] J. Zhang, B. Peng, S. Kim, F. Monifi, X. Jiang, Y. Li, P. Yu, L. Liu, Y.-x. Liu, A. Alù, Optomechanical dissipative solitons, Nature, 600 (2021) 75-80.

[8] J. Szabados, D.N. Puzyrev, Y. Minet, L. Reis, K. Buse, A. Villois, D.V. Skryabin, I. Breunig, Frequency comb generation via cascaded second-order nonlinearities in microresonators, Physical Review Letters, 124 (2020) 203902.
[9] A. Ganesan, C. Do, A. Seshia, Phononic frequency comb via intrinsic three-wave mixing, Physical review letters, 118 (2017) 033903.
[10] X. Zhang, B. Zhang, S. Wei, H. Li, J. Liao, C. Li, G. Deng, Y. Wang, H. Song, L. You, Telecom-band–integrated multimode photonic quantum memory, Science Advances, 9 (2023) eadf4587.
[11] Q. Guo, X.-Z. Qi, L. Zhang, M. Gao, S. Hu, W. Zhou, W. Zang, X. Zhao, J. Wang, B. Yan, Ultrathin quantum light source with van der Waals $NbOCl_2$ crystal, Nature, 613 (2023) 53-59.
[12] Z.L. Newman, V. Maurice, T. Drake, J.R. Stone, T.C. Briles, D.T. Spencer, C. Fredrick, Q. Li, D. Westly, B.R. Ilic, Architecture for the photonic integration of an optical atomic clock, Optica, 6 (2019) 680-685.
[13] P. Del'Haye, K. Beha, S.B. Papp, S.A. Diddams, Self-injection locking and phase-locked states in microresonator-based optical frequency combs, Physical review letters, 112 (2014) 043905.
[14] J. Leuthold, C. Koos, W. Freude, Nonlinear silicon photonics, Nature photonics, 4 (2010) 535-544.
[15] J. Lu, J.B. Surya, X. Liu, A.W. Bruch, Z. Gong, Y. Xu, H.X. Tang, Periodically poled thin-film lithium niobate microring resonators with a second-harmonic generation efficiency of 250,000%/W, Optica, 6 (2019) 1455-1460.
[16] J. Lin, N. Yao, Z. Hao, J. Zhang, W. Mao, M. Wang, W. Chu, R. Wu, Z. Fang, L. Qiao, Broadband quasi-phase-matched harmonic generation in an on-chip monocrystalline lithium niobate microdisk resonator, Physical review letters, 122 (2019) 173903.
[17] X. Guo, C.-L. Zou, H.X. Tang, Second-harmonic generation in aluminum nitride microrings with 2500%/W conversion efficiency, Optica, 3 (2016) 1126-1131.
[18] L. Chang, A. Boes, P. Pintus, J.D. Peters, M. Kennedy, X.-W. Guo, N. Volet, S.-P. Yu, S.B. Papp, J.E. Bowers, Strong frequency conversion in heterogeneously integrated GaAs resonators, APL Photonics, 4 (2019).
[19] J.-Q. Wang, Y.-H. Yang, M. Li, X.-X. Hu, J.B. Surya, X.-B. Xu, C.-H. Dong, G.-C. Guo, H.X. Tang, C.-L. Zou, Efficient frequency conversion in a degenerate $\chi^{(2)}$ microresonator, Physical Review Letters, 126 (2021) 133601.
[20] D.M. Lukin, C. Dory, M.A. Guidry, K.Y. Yang, S.D. Mishra, R. Trivedi, M. Radulaski, S. Sun, D. Vercruysse, G.H. Ahn, 4H-silicon-carbide-on-insulator for integrated quantum and nonlinear photonics, Nature Photonics, 14 (2020) 330-334.
[21] C. Wang, Z. Fang, A. Yi, B. Yang, Z. Wang, L. Zhou, C. Shen, Y. Zhu, Y. Zhou, R. Bao, High-Q microresonators on 4H-silicon-carbide-on-insulator platform for nonlinear photonics, Light: Science & Applications, 10 (2021) 139.
[22] J. Liu, F. Bo, L. Chang, C.-H. Dong, X. Ou, B. Regan, X. Shen, Q. Song, B. Yao, W. Zhang, Emerging material platforms for integrated microcavity photonics, Science China Physics, Mechanics & Astronomy, 65 (2022) 104201.
[23] R.S. Jacobsen, K.N. Andersen, P.I. Borel, J. Fage-Pedersen, L.H. Frandsen, O. Hansen, M. Kristensen, A.V. Lavrinenko, G. Moulin, H. Ou, Strained silicon as a new electro-optic material, Nature, 441 (2006) 199-202.
[24] M. Cazzanelli, F. Bianco, E. Borga, G. Pucker, M. Ghulinyan, E. Degoli, E. Luppi, V. Véniard, S. Ossicini, D. Modotto, Second-harmonic generation in silicon waveguides strained by silicon nitride, Nature materials, 11 (2012) 148-154.
[25] E. Timurdogan, C.V. Poulton, M. Byrd, M. Watts, Electric field-induced second-order nonlinear optical effects in silicon waveguides, Nature Photonics, 11 (2017) 200-206.
[26] X. Zhang, Q.-T. Cao, Z. Wang, Y.-x. Liu, C.-W. Qiu, L. Yang, Q. Gong, Y.-F. Xiao, Symmetry-breaking-induced nonlinear optics at a microcavity surface, Nature Photonics, 13 (2019) 21-24.
[27] J.S. Levy, M.A. Foster, A.L. Gaeta, M. Lipson, Harmonic generation in silicon nitride ring resonators, Optics express, 19 (2011) 11415-11421.

[28] X. Lu, G. Moille, A. Rao, D.A. Westly, K. Srinivasan, Efficient photoinduced second-harmonic generation in silicon nitride photonics, Nature Photonics, 15 (2021) 131-136.
[29] E. Nitiss, J. Hu, A. Stroganov, C.-S. Brès, Optically reconfigurable quasi-phase-matching in silicon nitride microresonators, Nature Photonics, 16 (2022) 134-141.
[30] Y. Xu, M. Han, A. Wang, Z. Liu, J.R. Heflin, Second order parametric processes in nonlinear silica microspheres, Physical review letters, 100 (2008) 163905.
[31] G. Kozyreff, J.L. Dominguez-Juarez, J. Martorell, Nonlinear optics in spheres: from second harmonic scattering to quasi-phase matched generation in whispering gallery modes, Laser & Photonics Reviews, 5 (2011) 737-749.
[32] See Supplemental Material at [URL will be inserted by publisher] for details of the theoretical models, numerical simulations and experimental details., DOI.
[33] N. Mukherjee, R. Myers, S. Brueck, Dynamics of second-harmonic generation in fused silica, JOSA B, 11 (1994) 665-669.
[34] A.W. Bruch, X. Liu, Z. Gong, J.B. Surya, M. Li, C.-L. Zou, H.X. Tang, Pockels soliton microcomb, Nature Photonics, 15 (2021) 21-27.

*Supplementary Materials for:*

# Molecule-induced surface second-order nonlinearity in an inversion symmetric microcavity

Ru Wang[#], Yue Dai[#], Jinsong Cheng, Ruoyu Wang, Xiaoqin Shen*

School of Physical Science and Technology, ShanghaiTech University, Shanghai, China 201210

[#] The authors contributed equally to this work.

Corresponding email: shenxq@shanghaitech.edu.cn

**Contents:** the supplementary information is organized into five sections and references.

**Section 1. The surface molecule contribution to the second-order nonlinearity**

**1.1 Surface SHG in the conventional free space laser platform**

The SHG in an inversion symmetry medium originates from both the bulk electric-quadrupole and surface electric-dipole responses, though the bulk electric-dipole response vanishes in the electric-dipole approximation. Given the surface layer is much less than a wavelength in thickness, the phenomenological model for the overall second-order nonlinear electric polarization $\boldsymbol{P}^2$ is typically expressed as[1]:

$$\boldsymbol{P}^{(2)} = 2\chi_q^{(2)}:\boldsymbol{E_1}\nabla\boldsymbol{E_1} - \nabla|\chi_Q^{(2)}:\boldsymbol{E_1}\boldsymbol{E_1}| + \chi_{sd}^{(2)}:\boldsymbol{E_1}\boldsymbol{E_1}, \tag{1}$$

where $\chi_q^{(2)}$ and $\chi_Q^{(2)}$ are the electric-quadrupole susceptibility in the bulk and at the surface, respectively; $\chi_{sd}^{(2)}$ is the electric-dipole susceptibility at the surface, and $\boldsymbol{E_1}$ is electric field of the pump. The first term denotes the contribution of electric-quadrupole response in bulk. The second term represents contribution of electric-quadrupole response at the surface/interface when an abrupt field exist due to the dielectric constant change. The third term indicates contribution of electric-dipole response at the surface/interface.

For air-fused silica (bare) interface, contribution from the bond electric-dipole response (the third term) is much less than 25% of the electric-quadrupole response (third first and second term)[2], due to the extreme weak bond electric-dipole. However, if an organic molecular monolayer absorbs on the silica surface, owing to the very large molecular polarizability, the molecular electric-dipole response (the third term) becomes a dominating contribution to the nonlinear polarization[3]. Based on it, surface SHG had been developed to be a popular nonlinear optical spectroscopic and microscopic technology, with sub-monolayer sensitivity[4].

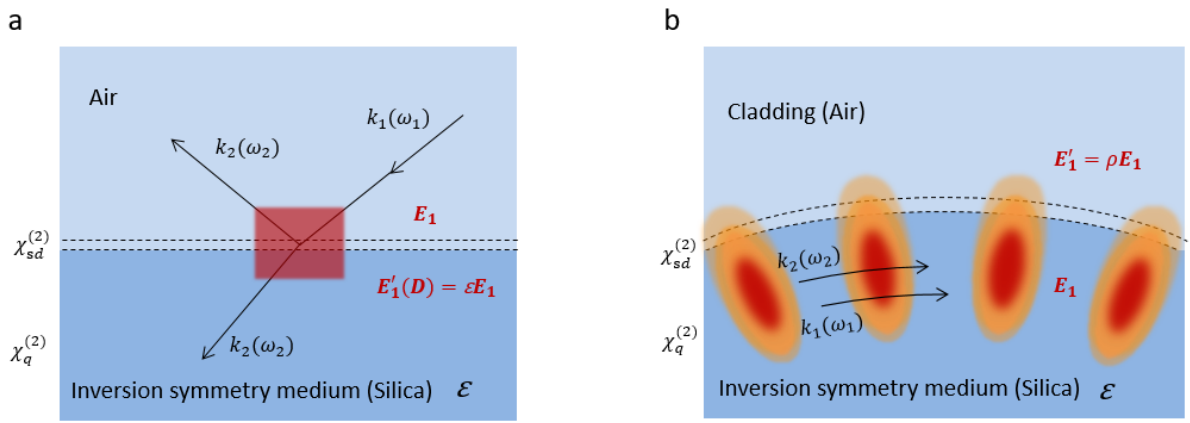

**Figure S1**. Schematic representation of phenomenological models for SHG at the interface/surface between two centrosymmetric media, air and fused silica. (a) Surface SHG in the conventional free space laser platform. The optical fields of the excitation light ($\omega_1$) on either side of the flat surface layer are identical

under the free space excitation. (b) Surface SHG in a whispering gallery mode microcavity. The electric field of the pump ($\omega_1$) in air (evanescent field) is much smaller that the peak electric field of the pump in silica ($E'_1 = \rho E_1, \rho \ll 1$).

However, it should be noted that the equation **(1)** is only valid with the assumption that the volume elements are small compared to the field variation length[1]. That is, the optical fields continuously across the interface are identical on either side of the interface, which is applicable to the conventional free space laser excitation on a flat surface of a transparent substrate (**Figure S1a**). The variation of electric field in the two medium arising from the difference in dielectric constants. For a whispering gallery mode microcavity, the phenomenological model should be modified to for the analysis of the surface second order nonlinear optical phenomenon.

### 1.2 Surface SHG in a whispering gallery mode microcavity

For a whispering gallery mode microcavity that the optical field is confined inside the cavity (as shown in **Figure S2**), the optical field beyond the surface/interface boundary (evanescent field) in the cladding (air) is much small than the field inside the cavity ($\rho = E'_1/E_1, \rho \ll 1$). In this case, the phenomenological model **Equation (1)** is no more applicable. Thus, we modify the expression to be:

$$P^{(2)} = 2\chi_q^{(2)}:E_1\nabla E_1 - \rho\chi_Q^{(2)}:E_1\nabla E_1 + \rho^2\chi_{sd}^{(2)}:E_1 E_1, \tag{2}$$

where the first term denotes the electric-quadrupole contribution due to the electric field gradient in the bulk; the second term represents the electric-quadrupole contribution from the rapid variation of the electric field along radical direction at the surface and from a longitudinal source electric-polarization at the surface; the third term indicates contribution of electric-dipole response at the surface/interface.

In the case of an inversion symmetry microcavity (silica, in our work) functionalized with a monolayer of molecules, $\chi_{sd}^{(2)}$ originates from the self-aligned monolayer of molecules, and thus can be replaced with $\chi_{mol}^{(2)}$. The typical $\chi_{mol}^{(2)}$ values of nonlinear molecules[5] are in the range of ~$10^1$ to ~$10^2$ pm/V, which are of about four to five orders of magnitude higher than the $\chi^2$ value of silica (~$10^{-3}$ pm/V)[6]. The overall second-order nonlinear electric polarization can be simplified as:

$$P^{(2)} = \chi_{silica}^{(2)}:E_1 E_1 + \rho^2\chi_{mol}^{(2)}:E_1 E_1, \quad \text{(functionalized silica microcavity)} \tag{5}$$

Thus, the effective $\chi^{(2)}$ for a surface molecule-functionalized silica microcavity can be expressed as:

$$\chi_{eff}^{(2)} = \chi_{silica}^{(2)} + \rho^2\chi_{mol}^{(2)}. \tag{6}$$

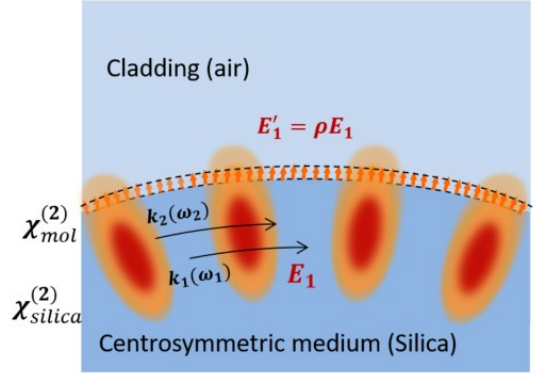

**Figure S2.** Schematic of the phenomenological model (equation 5) for SHG at the surface of a whispering gallery mode silica microcavity.

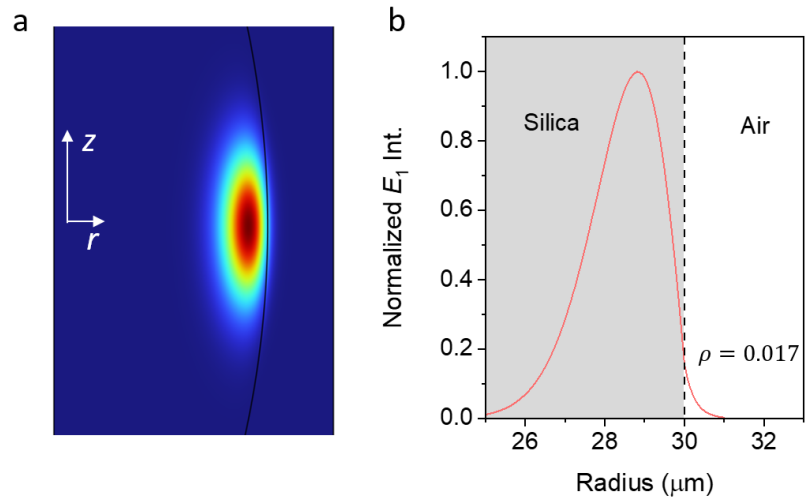

**Figure S3.** (a) FEM simulation of the optical mode profile in the microcavity with a radius of 30 μm, at 1550 nm. $z$ is the polar axis and $r$ is the radical axis. The black line indicates the boundary between the silica device and the cladding (air). (b) The mode profile at the cross section in the radical direction. The black dash line indicates the boundary. The optical field extended in air is about 1.7% of the total optical field ($\rho = 0.017$).

**Section 2. Surface molecule enhanced second-order nonlinear coupling strength**

The second-harmonic generation (SHG) process in a spherical microcavity involves two optical modes, and can be described by the Hamiltonian.

$$H = \omega_1 a_1^\dagger a_1 + \omega_2 a_2^\dagger a_2 + g\left[\left(a_1^\dagger\right)^2 a_2 + (a_2^\dagger)^2 a_1\right] + \epsilon_p(a_1 e^{i\omega_p t} + a_1^\dagger e^{-i\omega_p t}), \quad (6)$$

where the subscripts ($j$ = 1, 2) represent the pump and SH cavity modes with resonance frequency $\omega_1$ and $\omega_2$, respectively; the $\alpha_1$ and $\alpha_2$ are the Bosonic operators for the pump and SH modes in the WGM mocrocavity, respectively; $\epsilon_p = \sqrt{2\kappa_{1e} P_p / \hbar \omega_p}$ is the input pump filed strength, where $\kappa_{1e}$ is the external coupling rate of the pump mode and $P_p$ and $\omega_p$ are the power and frequency of the pump laser, respectively.

The overall nonlinear coupling strength $g$ between the pump and SH modes can be described as

$$g = \varepsilon_0 \iiint \frac{3\chi^{(2)}}{4\sqrt{2}\hbar} |E_1^*|^2 E_2 \sin\theta d\theta d\varphi, \quad (7)$$

where $\chi^{(2)}$ is the second-order susceptibility of microcavity medium, $\varepsilon_0$ is the vacuum permittivity, $E_j$ is the electric fields of the pump ($j$ = 1) and SH ($j$ = 2) mode, respectively.

The coupling strength can be obtained as

$$g \approx \chi^{(2)} \sqrt{\frac{\hbar \omega_1^2 \omega_2}{\varepsilon_0 2\pi R} \frac{\xi}{\varepsilon_1 \sqrt{\varepsilon_2}}} \frac{3}{4\sqrt{2}} \times \delta_{l,m,p}(l_2 - 2l_1), \quad (8)$$

where $\xi$ is the effective mode-overlapping factor at the cross-section of the microcavity; R is the radius of the miacrocavity; $l$, $m$ and $p$ are the orbital, azimuthal and radical quantum numbers, respectively. $g$ is non-zero only when $l_2 - 2l_1 = 0$, it ensures phase matching for WMG microcavities with $m = l$ (for both pump and SH modes).

In order to simply the discussion in the following sections, we simplified equation 8 as $g \approx K\chi^{(2)}$, where $K \equiv \sqrt{\frac{\hbar \omega_1^2 \omega_2}{\varepsilon_0 2\pi R} \frac{\xi}{\varepsilon_1 \sqrt{\varepsilon_2}}} \frac{3}{4\sqrt{2}} \times \delta_{l,m,p}(l_2 - 2l_1)$, is defined as the nonlinear coupling factor. $K$ involves three factors: the phase matching requirement, $l_2 - 2l_1 = 0$; the pump-SH mode overlapping factor, $\xi$; the radius of the microcavity, $R$. In the experiment, it is difficult to quantitatively analyze $K$. However, the $\chi^{(2)}$ values for wide range of different materials are well characterized and reported. For a surface molecule functionalized silica microcavity, the overall coupling strength can be expressed as:

$$g \approx K\left(\chi_{silica}^{(2)} + \rho^2 \chi_{mol}^{(2)}\right). \quad (9)$$

**Section 3. Coupled mode equation for second harmonic generation**

According to Heisenberg equation, the coupled-mode equations satisfy

$$\frac{d\alpha_1}{dt} = [-i(\omega_1 - \omega_P) - \kappa_1]\alpha_1 - i2g\alpha_1^*\alpha_2 - i\epsilon_p, \tag{10}$$

$$\frac{d\alpha_2}{dt} = [-i(\omega_2 - 2\omega_P) - \kappa_2]\alpha_2 - ig\alpha_1^2. \tag{11}$$

By incorporating the loss and input, the above coupled-mode equations are given

$$\frac{d\alpha_1}{dt} = \left[-i(\omega_1 - \omega_P) - \frac{\kappa_{10} + \kappa_{1e}}{2}\right]\alpha_1 - ig\alpha_1^*\alpha_2 + \sqrt{\kappa_{1e}}s, \tag{12}$$

$$\frac{d\alpha_2}{dt} = \left[-i(\omega_2 - 2\omega_P) - \frac{\kappa_{20} + \kappa_{2e}}{2}\right]\alpha_2 - ig\alpha_1^2, \tag{13}$$

where $s$ is the amplitude of the input electric field in the pump fiber with $|s|^2$ being the input power; $\kappa_{j0}$ and $\kappa_{je}$ denote the intrinsic decay rate and external coupling rate, respectively.

Because the field in the pump cavity mode is not affected by the SHG, the second term in (S13) can be ignored. Considering the steady state condition, the output power of SH mode can be obtained,

$$P_2 = \frac{4Q_2^2/Q_{2e}}{4Q_2^2/\omega_2(2\omega_p - \omega_2) + \omega_2}\left(\frac{4Q_1^2/Q_{1e}}{4Q_1^2/\omega_1(\omega_p - \omega_1) + \omega_1}\right)^2 g^2 P_1^2, \tag{14}$$

where $Q_j = \omega_j/(\kappa_{j0} + \kappa_{je})$ and $Q_{je} = \omega_j/\kappa_{je}$ represent the loaded quality factor and the external quality factor, respectively.

By incorporating (S10) to (S15), SHG efficiency, as can be expressed as

$$P_2/P_1^2 = \frac{4Q_2^2/Q_{2e}}{4Q_2^2/\omega_2(2\omega_p - \omega_2) + \omega_2}\left(\frac{4Q_1^2/Q_{1e}}{4Q_1^2/\omega_1(\omega_p - \omega_1) + \omega_1}\right)^2 K^2\left[\chi_{silica}^{(2)} + \rho^2\chi_{mol}^{(2)}\right]^2. \tag{15}$$

Under dual mode resonance phase matching condition (i.e., $2\omega_p = \omega_2 = \omega_1$) and critical coupling (i.e., $Q_{j0} = Q_{je}$ and $Q_j = 2Q_{je}$), it can be simplified as

$$P_2/P_1^2 = \frac{4Q_2}{\omega_2}\left(\frac{4Q_1}{\omega_1}\right)^2 K^2\left[\chi_{silica}^{(2)} + \rho^2\chi_{mol}^{(2)}\right]^2. \tag{16}$$

The above equation clearly describes the dependency of the SH efficiency on the molecular surface second-order nonlinearity. It is the same as equation 1 in the main text.

**Section 4. Surface molecule-induced SH enhancement**

From equation 17, we can obtain the contribution of surface molecular nonlinearity to the enhancement of SHG. Given the Q values and $K$ remain constant before and after surface molecule functionalization, the surface-molecule-induced SH enhancement factor $\eta_{mol}$ can be expressed as:

$$\eta_{mol} = \left(1 + \rho^2 \chi_{mol}^{(2)}/\chi_{silica}^{(2)}\right)^2, \quad (theoritical) \tag{17}$$

Experimentally, the $\eta_{mol}$ value can be retrieved from the "in situ" experimental result. In the "in situ" experiment, $K$ remains constant for a same pump mode and the Q values (assuming $Q_1 = Q_2$) before and after surface functionalization can be measured experimentally. With a fixed pump power and under dual resonance phase matching and critical coupling conditions, the experimental $\eta_{mol}$ value can be calculated from the following expression:

$$\eta_{mol} = \frac{P_{2,before}}{P_{2,after}} \left(\frac{Q_{after}}{Q_{before}}\right)^3, \quad (experimental) \tag{18}$$

where subscripts *before* and *after* represent the microscavity before and after surface functionalization in the "in situ" experiment, respectively. Based on the experimental data from Figure 5 in the main text, under TM pump, the $\eta_{mol}$ is retrieved to be about 216. This results agrees well with the theoretical value of 260 calculated from equation 18, for the functionalized silica microcavity with $\rho = 0.017$ and $\chi_{mol,\perp}^{(2)} = 210$ pm/V and $\chi_{silica}^{(2)} = 4 \times 10^{-3}$ pm/V.

From equation 17, it can be found that, not only the surface molecular nonlinear susceptibility $\chi_{mol}^{(2)}$ but also the surface evanescent field is deterministic to the surface-molecule-induced SH enhancement factor. As shown in **Figure S3**, the enhancement factor would be further improved to four-orders of magnitude by increasing the $\rho$ value to 0.05. Moreover, from **Figure S4** it can be found that that the maximum SH enhancement factor can be as high as about nine to ten orders of magnitude, in the situation that a molecular monolayer sheet is embed inside the mirocavity structure to interact with the peak optical field ($\rho = 1$).

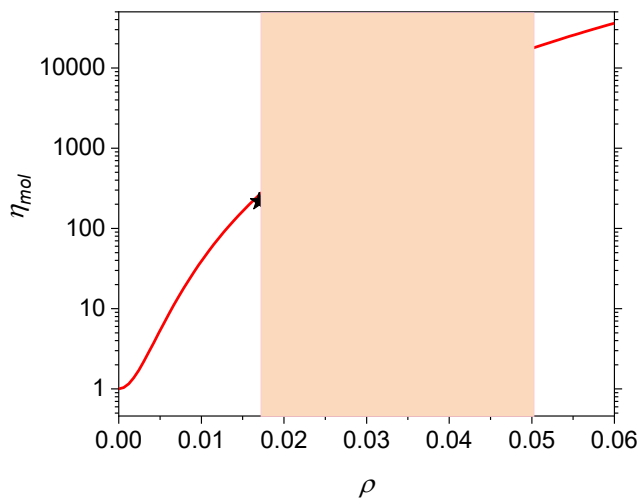

**Figure S3. The dependency of surface molecule induced SH enhancement factor on the surface evanescent field, as described in equation 18.** It indicates that, by increase the $\rho$ value from 0.17 to 0.05 via proper geometry design, the SH enhancement factor can be further increase by two-orders of magnitude, up to about four-orders of magnitude. The back star is the experimental data for a silica microsphere with $\rho = 0.017$.

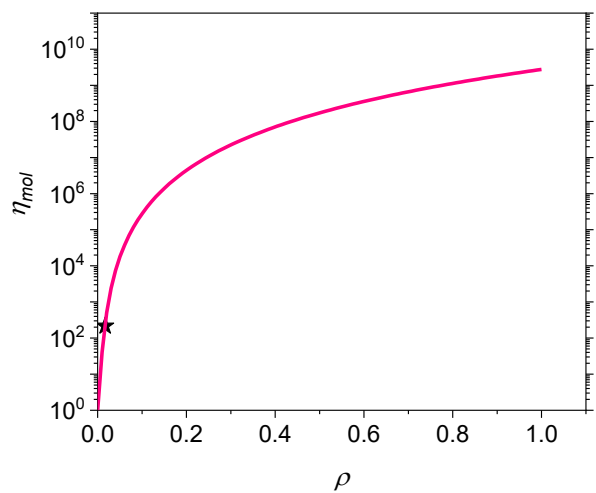

**Figure S4. The theoritical capability of the molecule-induced SH enhancement.** It reveals that the maximum SH enhancement factor can be as high as about nine to ten orders of magnitude, in the situation that an asymmetric molecular monolayer sheet is embed inside the mirocavity structure interacting with the peak optical field ($\rho = 1$).

**Section 5. Device fabrication and surface characterization**

The silica microsphere cavities are fabricated from a standard single-mode silica fiber SMF-28. The optical fiber is tapered and then melted to form a microsphere with the pulsed CO2 laser. The surface functionalization of the silica microspheres are conducted using modified method[7]. Briefly, as-prepared silica microspheres are firstly treated by O2 plasma to generate dense hydroxyl groups on the silica surface. Then, the microspheres are grafted with a coupling layer by using [4-(chloromethyl)phenyl]trichlorosilane, followed by drop-casting with a layer of the 4-[4-diethylamino(styryl)]pyridinium molecules from a stored solution in THF with different concentrations (~$1 \times 10^{-3}$ to $1 \times 10^{-9}$ mol/L). The drop-casted devices are incubated at 120°C under vacuum for about 20 minutes to accelerate the quaternization reaction. The microspheres are then cooled to room temperature and rinsed thoroughly with tetrahydrofuran and dried under vacuum at 100°C for 5 minutes, yielding a surface functionalized microsphere.

To characterize the chemical components of the surface molecular layer, silicon wafers with thermo-oxide are used to prepare a molecule-factionalized silica surface with the same processes, as aforementioned. The surface chemical components of the silica devices characterized by X-ray photoelectron spectroscopy (ESCALAB 250XI, ThermoFisher) to optimize the process to obtain the functionalized silica devices with a monolayer of grafted molecules. The high-resolution N 1s XPS spectrum of the optimized stibazolium molecule-functionalized silica surface shows a broad peak centered at ~ 400 eV, which can be de-convoluted into two peaks with binding energies of ~399.4 eV and ~402.0 eV for the amino and pyridinium nitrogen atoms, respectively. It confirms that the stibazoilum molecules are covalent-bonded on the surface, resulting in an asymmetric alignment of molecules. For the samples prepared from the store solution with concentration above $1 \times 10^{-4}$ mol/L, the high-resolution N 1s XPS spectra exhibit a binding energy peak at ~ 400.4 eV corresponding to pyridine nitrogen atom. It indicates that the existence of free 4-[4-diethylamino(styryl)]pyridinium molecules on the surface through intermolecular dipole-dipole interaction with the grafted stibazoilum molecules. With the store solution with concentration bellow $1 \times 10^{-6}$ mol/L, the binding energy peak at ~ 400.4 eV disappear, suggesting that no physical absorption of free molecules occur. The absence of physical absorption is of critical for the second-order nonlinear enhancement, because such intermolecular interaction would cancel the asymmetry of the functionalized molecular layer.

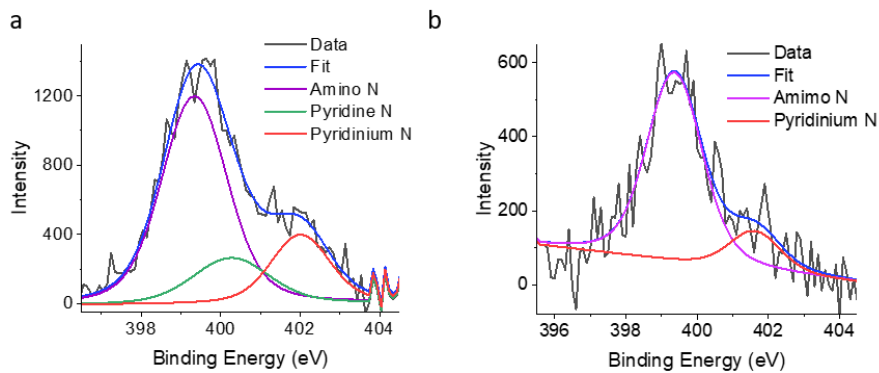

**Figure S5. XPS spectrum of the surface.** The high-resolution N 1s XPS spectra of samples prepared from store solution with concentration of ~1 × 10$^{-4}$ mol/L (a) and of ~1 × 10$^{-6}$ mol/L (b), respectively.

To characterize the thickness of the surface molecular monolayer, a silicon wafer with thermo-oxide is first patterned with photoresist by a standard photolithography method to define the molecular functionalization areas. The patterned silicon wafer is then grafted with a layer of stibazoilum molecules with the same process as mention above. After that, the remaining photoresist is removed completely by using acetone, methanol, and isopropanol rinsing cycle followed by drying with a nitrogen gas air gun to yield a patterned surface stibazoilum molecular layer with clear boundary for mapping the step heights of the surface by using an atomic force microscope (Cypher S, Oxford/Asylum Research)

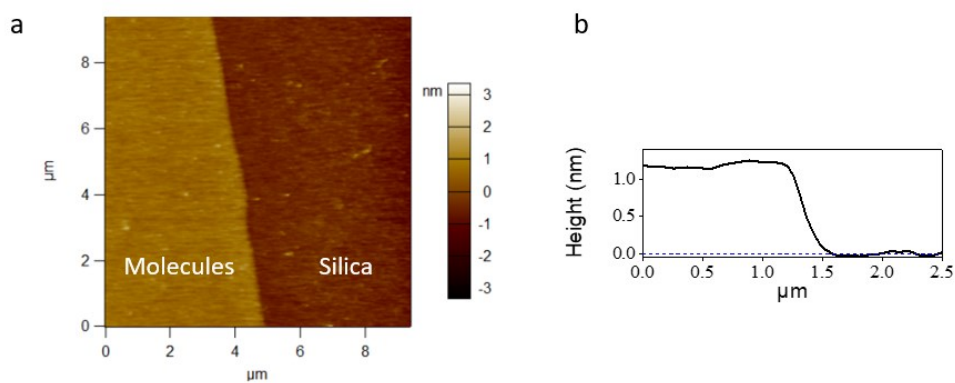

**Figure S6. AFM measurement of the surface. a**, AFM mapping of a molecular functionalized surface. **b**, The measured step height of the functionalized molecular layer across the boundary in (**a**). The measured thickness of the stibazoilum monolayer is of about 1.2 nm, which is consistent with the height of a monolayer of the asymmetrically aligned molecules with an average tilt angle of about 40°. Thus $\chi^{(2)}_{mol,\perp}$ can be calculated to be ~ 210 pm/V based on $\chi^{(2)}_{mol,\perp} = \chi^{(2)}_{mol} cos\theta$, where $\chi^{(2)}_{mol} = $ ~280 pm/V is obtained from the literature[8].

# References

1   Ponath, H.-E. & Stegeman, G. I. *Nonlinear surface electromagnetic phenomena*. (Elsevier, 2012).
2   Guyot-Sionnest, P. & Shen, Y. Local and nonlocal surface nonlinearities for surface optical second-harmonic generation. *Physical Review B* **35**, 4420 (1987).
3   Chen, C., Heinz, T., Ricard, D. & Shen, Y. Detection of molecular monolayers by optical second-harmonic generation. *Physical Review Letters* **46**, 1010 (1981).
4   Sun, S., Tian, C. & Shen, Y. R. Surface sum-frequency vibrational spectroscopy of nonpolar media. *Proceedings of the National Academy of Sciences* **112**, 5883-5887 (2015).
5   Munn, R. W. & Ironside, C. N. *Principles and applications of nonlinear optical materials*. (Springer, 1993).
6   Mukherjee, N., Myers, R. & Brueck, S. Dynamics of second-harmonic generation in fused silica. *JOSA B* **11**, 665-669 (1994).
7   Shen, X., Beltran, R. C., Diep, V. M., Soltani, S. & Armani, A. M. Low-threshold parametric oscillation in organically modified microcavities. *Science advances* **4**, eaao4507 (2018).
8   Lin, W. *et al.* New Nonlinear Optical Materials: Expedient Topotactic Self‐Assembly of Acentric Chromophoric Superlattices. *Angewandte Chemie International Edition in English* **34**, 1497-1499 (1995).